# Maximum Likelihood Associative Memories

Vincent Gripon
Electronics Department
Télécom Bretagne, Brest, France
vincent.gripon@ens-cachan.org

Michael Rabbat
Department of Electrical and Computer Engineering
McGill University, Montréal, Canada
michael.rabbat@mcgill.ca

*Abstract*—Associative memories are structures that store data in such a way that it can later be retrieved given only a part of its content—a sort-of error/erasure-resilience property. They are used in applications ranging from caches and memory management in CPUs to database engines. In this work we study associative memories built on the maximum likelihood principle. We derive minimum residual error rates when the data stored comes from a uniform binary source. Second, we determine the minimum amount of memory required to store the same data. Finally, we bound the computational complexity for message retrieval. We then compare these bounds with two existing associative memory architectures: the celebrated Hopfield neural networks and a neural network architecture introduced more recently by Gripon and Berrou.

## I. Introduction

In this work, we focus on a specific kind of memory termed "associative". Contrary to classical index-based memories, associative ones are able to retrieve a piece of information given only a part of it. In that sense, they perform similarly to human memories that are able, for instance, to remember that Claude Shannon wrote "A Mathematical Theory of Communication" without having to list or search through all authors with whom they are familiar.

Associative memories are used in a variety of applications including caches for processing units [1], databases engines [2], and intrusion prevention systems [3]. Associative memories typically provide an attractive balance between speed of lookup and memory cost, often at an increase in retrieval error rate.

A trivial way to implement an associative memory is to use a brute-force approach that tests each message in the stored dataset to find a match. Because it is possible to evaluate these tests concurrently, this strategy is often used in electronics [4], at the cost of high power consumption.

Other approaches include neural network-based architectures. A typical example is the Hopfield Neural Network that was introduced by J.J. Hopfield in 1982 [5]. This work inspired many scientists in communities from machine learning [6] to statistical physics [7]. Other examples of associative memories include Kohonen maps [8] and Boltzmann machines [9].

Recently, Gripon and Berrou introduced a novel architecture [10], [11] for associative memories based on neural networks. Their architecture outperforms Hopfield networks in terms of diversity (number of words stored) and capacity (total number of bits stored) for the same memory used.

In this work, we study the performance of an associative memory based on the maximum likelihood principle. We present bounds on the error rate, memory complexity, and computational complexity of this associative memory, and then we compare these bounds with the models proposed by Hopfield and by Gripon and Berrou, which can be viewed as approximations to the ML associative memory.

The structure of the paper is as follows. Section II presents previous work on the topic. Section III gives the definitions and formally introduces the ML associative memory. Section IV presents the models previously discussed. Section V gives bounds on error rate when storing random sets of words. Section VI and VII are dedicated respectively to the memory and computational complexity, and we conclude in Section VIII.

## II. Previous work

McEliece et al. [12] introduce bounds on Hopfield neural networks. They use techniques from coding theory to obtain the maximum number of messages a Hopfield network with $n$ neurons can store and correctly retrieve when the Hamming distance between the initial codeword and that presented to the device is less than $n/2$.

Yaakobi and Bruck [13] also study bounds for associative memories. They consider a Hamming space, and search for the minimum number of spheres one must intersect in order to find an element in the space. They consider a definition of associative memory, where associated elements are those with small Hamming distance, which is fundamentally different from the definition considered in this paper.

Karbasi, Salavati, and Shokrollahi [14], [15] study multi-layer neural architectures which are capable of storing large sets of messages under the assumption that the stored messages satisfy a redundancy (low rank) condition.

Gripon, Rabbat, Skachek and Gross [16] study the efficient storage of unordered sets and introduce minimum bounds on memory size based on entropy. The results we present in Section VI use similar arguments.

## III. Problem Formulation

Let $\mathcal{A}$ be a finite alphabet with $|\mathcal{A}| \geq 2$, and consider a set $\mathcal{S}$ of $m$ words in $\mathcal{A}^n$. For a word $w \in \mathcal{A}^n$ of length $n$, we denote by $w(i)$ the $i$-th symbol of $w$. Provide $\mathcal{S}$ with a probability measure $\mu$, and consider a random function:

$$e : \begin{array}{ccc} \mathcal{S} & \to & \overline{\mathcal{A}}^n \\ w & \mapsto & e(w) \end{array}$$

where $\overline{\mathcal{A}} = \mathcal{A} \bigsqcup \{\bot\}$ and $\bot$ denotes an erased symbol.

**Definition 1.** An *associative memory* is an algorithm which, given a set $\mathcal{S} \subset \mathcal{A}^n$, learns a retrieval rule $f : \overline{\mathcal{A}}^n \to \mathcal{S}$.

Associative memories clearly have similarities to coding for the erasure channel. However, rather than jointly designing a code and decoding rule, an associative memory must be able to provide a decoder $f$ for any given subset $\mathcal{S} \subset \mathcal{A}^n$ of messages.

Let $\mathcal{F}(\mathcal{S})$ denote the set of all mappings from $\overline{\mathcal{A}}^n$ to $\mathcal{S}$.

**Definition 2.** The *success probability* of an associative memory $f \in \mathcal{F}(\mathcal{S})$ is

$$P_\mathcal{S}(f) = \sum_{w \in \mathcal{S}} \mu(w) \cdot P(f(e(w)) = w),$$

where $P(f(e(w)) = w)$ is the probability that the image of $e(w)$ under $f$ is $w$, under the distribution governing $e$.

It is well-known that the maximum likelihood decoding principle minimizes the error rate.

**Definition 3.** A *maximum likelihood associative memory* (ML-AM) is an algorithm which, given set of messages $\mathcal{S}$, returns a mapping $f^*$ which maximizes $P_\mathcal{S}(f)$ over all $f \in \mathcal{F}(\mathcal{S})$. The residual error rate of the ML-AM is

$$P^*_{\text{err}}(\mathcal{S}) = 1 - P_\mathcal{S}(f^*).$$

In general, an associative memory is designed to store arbitrary sets $\mathcal{S} \subset \mathcal{A}^n$, and thus the ML-AM is unlikely to have zero residual error. For instance, take $\mathcal{S}$ to be the set of four-letter words in the English language, and consider the task of retrieving a word given the input "hea$\bot$". Possible solutions include "head", "heap", and "heat", and thus $P^*_{\text{err}}(\mathcal{S}) \neq 0$.

## IV. HOPFIELD AND GRIPON-BERROU NEURAL NETWORKS

Hopfield neural networks (HNNs) [5] and Gripon-Berrou neural networks (GBNNs) [10] are two architectures that implement approximations to ML-AMs. Each architecture is defined by three items: 1) the network structure, 2) a rule for storing messages in the set $\mathcal{S}$, and 3) a rule for retrieving a message given an input in $\overline{\mathcal{A}}^n$.

HNNs store length-$n$ binary words over $\{-1, 1\}$ using a $n$-node complete graph. Information is encoded in integer-valued edge weights (i.e., potentials) which are stored in a symmetric matrix $(M_{ij})_{1 \leq i,j \leq n}$. Given $\mathcal{S}$, the weight on edge $(i,j)$ is

$$M_{ij} = \sum_{w \in \mathcal{S}} w(i) \cdot w(j).$$

Note that the weights can take up to $m+1$ distinct values, where $m$ is the size of $\mathcal{S}$.

When given an input $\overline{w} = e(w)$ with erased symbols, the matrix $M$ is used to try to retrieve the original word $w$. Denote by $\mathbf{v}_t$ the vector of values of nodes at time $t$, and initialize $\mathbf{v}_0 = \overline{w}^T$, replacing all instances of $\bot$ with 0. The algorithm repeats the iterations

$$\mathbf{v}_{t+1} = \text{sign}(M_{ij}\mathbf{v}_t),$$

where the $i$th coordinate of $\text{sign}(\mathbf{v})$ is 1 if $\mathbf{v}(i) \geq 0$ and $-1$ otherwise.

When the iterations above converge, the values at each node are output as the estimate of the original word. The problems of proof and speed of convergence of the algorithm are complex and have been studied [17], [18], [19]. In the remainder of this work, we use a lower bound of a finite number of iterations in order to allow comparison with GBNN.

GBNNs also use a lossy encoding of the set of words, but contrary to HNNs, this one is based on connection existence. The alphabet here can be any, but for more simplicity we suppose it to be the integers from 1 to $l$. In this model, the graph is binary and sparse. It is characterized by a number of divisions $n$, and a matrix $(W_{(i_1,j_1)(i_2,j_2)})_{0 \leq i_1,i_2 \leq n-1, 0 \leq j_1,j_2 \leq l-1}$. After storing $\mathcal{S}$, the matrix is defined as follows:

$$W_{(i_1,j_1)(i_2,j_2)} = \begin{cases} 1 & \text{if } \exists w \in \mathcal{S}, w(i_1) = j_1 \text{ and } w(i_2) = j_2 \\ 0 & \text{otherwise.} \end{cases}$$

To retrieve the initial word $w$ given an erased input $\overline{w}$, the algorithm also uses iterations to compute the value $\mathbf{v}_t$ of neurons at step $t$. The initialization is such that the $(i,j)$-th coordinate of $\mathbf{v}_0$ is 1 if $\overline{w}(i) = j$ and 0 otherwise. Then:

$$\mathbf{v}_{t+1} = s(W\mathbf{v}_t + \gamma \mathbf{v}_t) \quad (1)$$

where the $(i_0, j)$-th coordinate of $s(\mathbf{v})$ is 1 if its value is maximum among the coordinates $(i_0, j'), 1 \leq j' \leq l$, and it is 0 otherwise.

In practice, with $\gamma \geq nl$, it has been observed that a few iterations suffice to reach a fixed point. In particular, the error equations derived in [20] hold after a single iteration. Thus we suppose this number to be constant for the rest of this work.

Note that we do not compare with the memories of [14], [15] since they use real-valued edge potentials and thus cannot be exactly represented using a finite number of bits.

## V. RESIDUAL ERROR ON RANDOM SETS

We now consider finding the maximum number of codewords that can be stored in a ML-AM given a desired residual error probability. Let us focus on a simple case where the set $\mathcal{S}$ is drawn uniformly, and let us also assume that $\mu$ is the uniform distribution on $\mathcal{A}^n$.

**Theorem 4.** *Assume that $\mathcal{S}$ is drawn uniformly from all sets that contain $m$ words in $\mathcal{A}^n$, and suppose that $e$ erases exactly $r$ symbols of its argument at positions drawn uniformly and independently of $\mathcal{S}$. Then for the ML-AM mapping $f^*$, we have*

$$\mathbb{E}[P_\mathcal{S}(f^*)] = \frac{|\mathcal{A}|^{n-r}}{m} \left(1 - \frac{\binom{|\mathcal{A}|^n - m}{|\mathcal{A}|^r}}{\binom{|\mathcal{A}|^n}{|\mathcal{A}|^r}}\right), \quad (2)$$

*where the expectation is taken with respect to $\mathcal{S}$.*

*Proof:* First consider a particular realization of the set $\mathcal{S}$ and let $f^*$ denote the corresponding ML-AM retrieval rule.

Let

$$d(w_1, w_2) = \begin{cases} 0 & \text{if } \forall i, \vee \begin{array}{rcl} w_1(i) & = & w_2(i) \\ w_1(i) & = & \bot \\ w_2(i) & = & \bot \end{array} \\ 1 & \text{otherwise} \end{cases}, \quad (3)$$

be the function which indicates whether $w_1$ and $w_2$ are equal, up to erased symbols. Let $\overline{\mathcal{A}}_r^n$ denote the set of words $\overline{w}$ in $\overline{\mathcal{A}}^n$ that contain exactly $r$ symbols $\bot$, and let $\overline{S}_r = \{\overline{w} \in \overline{\mathcal{A}}_r^n : \exists w \in \mathcal{S}, d(w, \overline{w}) = 0\}$. Also let $S(\overline{w}) = \{w' \in \mathcal{S} : d(w', \overline{w}) = 0\}$, and note that $w \in S(e(w))$ if $w \in \mathcal{S}$. Without loss of generality, we restrict our analysis to mappings that always output a value in $S(\overline{w})$.

For the ML-AM retrieval rule, we have

$$\begin{aligned} m \cdot P_{\mathcal{S}}(f^*) &= \sum_{w \in \mathcal{S}} \sum_{\overline{w} \in \overline{\mathcal{A}}^n} P(f^*(\overline{w}) = w | \overline{w} = e(w)) \cdot P(\overline{w} = e(w)) \\ &= \binom{n}{r}^{-1} \sum_{\overline{w} \in \overline{S}_r} \sum_{w \in S(\overline{w})} P(f^*(\overline{w}) = w | \overline{w} = e(w)) \\ &= \binom{n}{r}^{-1} |\overline{S}_r| . \end{aligned}$$

In other words, given a partially erased word $\overline{w}$ as input, the best one can do is to choose any word in $e^{-1}(\overline{w})$. The selection process has no impact on performance.

The arguments above are for a particular set $\mathcal{S}$. To obtain the expected residual error rate of a ML-AM, let us find the expected value of $|\overline{S}_r|$ when the set of words to store is obtained by sampling $m$ words uniformly from $\mathcal{A}^n$. We first point out that there are $|\mathcal{A}|^r$ words in $\mathcal{A}^n$ that are at distance zero from a given word $\overline{w}$ in $\overline{\mathcal{A}}_r^n$. The probability that at least one of them is in $\overline{S}_r$ is

$$P(\overline{w} \in \overline{S}_r | \overline{w} \in \overline{\mathcal{A}}_r^n) = 1 - \frac{\binom{|\mathcal{A}|^n - m}{|\mathcal{A}|^r}}{\binom{|\mathcal{A}|^n}{|\mathcal{A}|^r}},$$

since this event follows a hypergeometric law. We obtain that

$$\mathbb{E}[|\overline{S}_r|] = |\mathcal{A}|^{n-r} \binom{n}{r} P(\overline{w} \in \overline{S}_r | \overline{w} \in \overline{\mathcal{A}}_r^n),$$

and thus

$$\mathbb{E}[P_{\mathcal{S}}(f^*)] = \frac{|\mathcal{A}|^{n-r}}{m} P(\overline{w} \in \overline{S}_r | \overline{w} \in \overline{\mathcal{A}}_r^n).$$

∎

*Remark* 5. If $m = o(|\mathcal{A}|^n - |\mathcal{A}|^r)$, then using Stirling's approximation we obtain that

$$\begin{aligned} 1 - \frac{\binom{|\mathcal{A}|^n - m}{|\mathcal{A}|^r}}{\binom{|\mathcal{A}|^n}{|\mathcal{A}|^r}} &= 1 - \frac{(|\mathcal{A}|^n - m)! \, (|\mathcal{A}|^n - |\mathcal{A}|^r)!}{(|\mathcal{A}|^n)! \, (|\mathcal{A}|^n - |\mathcal{A}|^r - m)!} \\ &\underset{n \to \infty}{\sim} 1 - \left(\frac{|\mathcal{A}|^n - |\mathcal{A}|^r}{|\mathcal{A}|^n}\right)^m \\ &\underset{n \to \infty}{\sim} 1 - \exp\left(-m|\mathcal{A}|^{r-n}\right), \end{aligned}$$

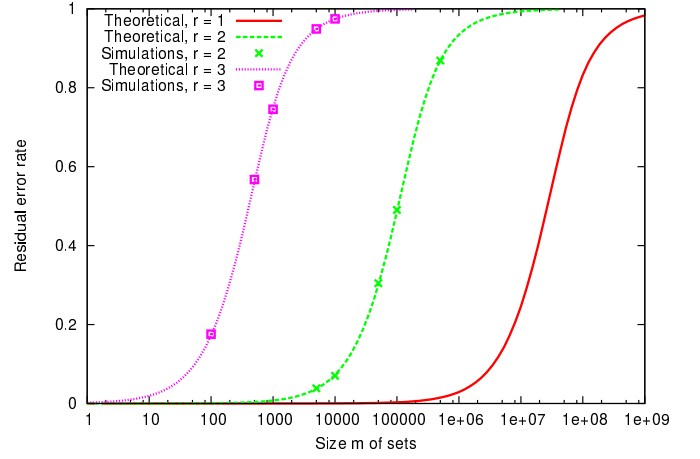

Fig. 1. Evolution of the ML-AM residual error rate as a function of $m = |\mathcal{S}|$, with alphabet size 256 and 4 symbols, and for various values of $r$.

and thus

$$\mathbb{E}[P_{\mathcal{S}}(f^*)] \underset{n \to \infty}{\sim} \frac{|\mathcal{A}|^{n-r}}{m} \left(1 - \exp(-m|\mathcal{A}|^{r-n})\right). \quad (4)$$

If, in addition, $r = o(n)$, then

$$\mathbb{E}[P_{\mathcal{S}}(f^*)] \underset{n \to \infty}{\approx} 1.$$

From Equation (4), one can estimate the number of words $m$ which can be stored while achieving a desired residual error probability $P_0$:

$$m \approx 2P_0 |\mathcal{A}|^{n-r}.$$

We note that if $r = \Theta(n)$, then the number of words grows exponentially with $n$ and polynomially with $|\mathcal{A}|$.

McEliece et al. [12] show that, under similar conditions, the number of words one can store in a HNN grows sub-linearly with $n$. For GBNNs, it is shown in [21] that the number of words grows quadratically with $n$.

Figure 1 depicts the evolution of the residual error rate $P^*_{\text{err}}(\mathcal{S})$ as a function of the size $m$ of $\mathcal{S}$ and for various values of $r$ when $|\mathcal{A}| = 256$ and $n = 4$.

HNNs and GBNNs are not ML-AMs and they may have a larger error probability. Figure 2 shows the evolution of the word retrieval error rate as a function of the number of erasures in input, together with the corresponding residual error.

## VI. MEMORY NEEDED

A ML-AM needs to record information about the stored set of words. Thus, the number of bits needed to represent a ML-AM is at least that required to losslessly encode the (unordered) set $\mathcal{S}$. In this section we estimate the number of bits that are required to implement a ML-AM. Similar developments for multisets have been proposed in [16].

Let us consider a set $\mathcal{S}$ as described in the previous section. The Kraft inequality [22] tells us that one cannot expect to use fewer than $H$ bits to represent the source, where $H$ is the entropy of the source generating these sets:

$$H = \log_2\left(\binom{|\mathcal{A}|^n}{m}\right).$$

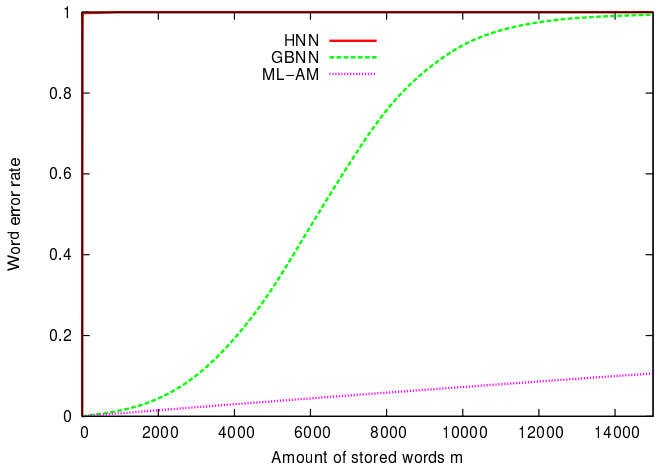

Fig. 2. Evolution of the word error rate when using HNN, GBNN (as described in [20] and with random decision on ambiguity) and a ML-AM, as a function of $m = |\mathcal{S}|$ and for $|\mathcal{A}| = 256$, $r = 2$ and $n = 4$.

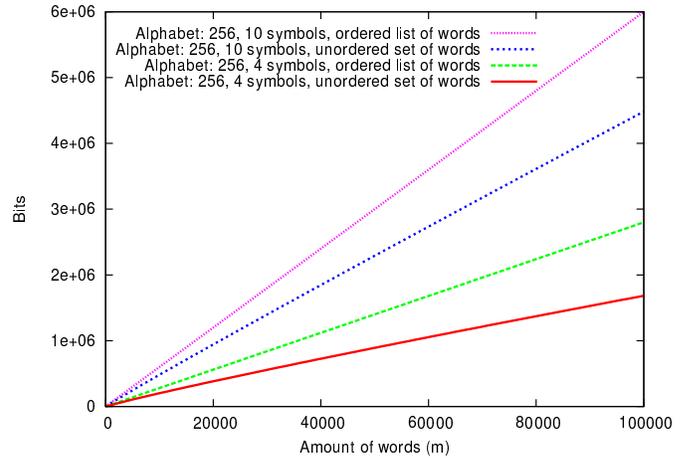

Fig. 3. Evolution of the number of bits required when representing random set of words of size $m$ and a corresponding ordered list of elements as functions of $m$ and for various alphabet size and number of symbols.

Using Stirling's approximation to estimate $H$ gives

$$H \underset{n,m\to\infty}{\sim} |\mathcal{A}|^n \log_2(|\mathcal{A}|^n) - m\log_2(m)$$
$$- (|\mathcal{A}|^n - m)\log_2(|\mathcal{A}|^n - m).$$

Suppose that $m = o(|\mathcal{A}|^n)$. Then

$$H \underset{n,m\to\infty}{\sim} m\log_2(|\mathcal{A}|^n),$$

which implies that if the number of words is small, it is equivalent (in terms of the number of bits) to store an ordered list of messages or the corresponding unordered set.

Let us consider another case where we denote by $c$ the ratio $|\mathcal{A}|^n/m$, and suppose that $c$ is constant as $m, n \to \infty$. Then

$$H \underset{n,m\to\infty}{\sim} m\left(c\log_2(c) - c\log_2(c-1) + \log_2(c-1)\right).$$

In this regime, the ratio of the expected number of bits to represent the unordered set to that of the corresponding ordered list of words is $(c\log_2(c) - c\log_2(c-1) + \log_2(c-1))/(n\log_2(|\mathcal{A}|))$. Note that this ratio tends to zero when $n$ tends to infinity.

Figure 3 depicts both the number of bits required to represent an unordered set $\mathcal{S}$ and that of a corresponding ordered list of elements as a function of $m$.

HNNs and GBNNs can only be reasonably used in the case of $m = o(|\mathcal{A}|^n)$. In this case, Table I compares the memory needed for both systems when targeting a residual error probability of 1/100. Note that for both HNNs and GBNNs the amount of memory needed could be reduced via standard compression techniques since the distribution of connection weights for HNNs and the distribution of connections for GBNN are not uniform. This reduction would nevertheless imply a larger computational complexity.

TABLE I
RATIO OF THE MEMORY NEEDED BY HNN AND GBNN TO THE ENTROPY $H$ FOR TARGET ERROR RATE $P_0 = 0.01$ AND FOR VARIOUS VALUES OF $|\mathcal{A}|, r,$ AND $n$. THESE NUMBERS WERE OBTAINED THROUGH SIMULATION.

| $|\mathcal{A}|$ | $n$ | $r$ | HNN | GBNN |
|---|---|---|---|---|
| 256 | 4 | 1 | 1619% | 684% |
| 64 | 10 | 1 | 3760% | 186% |
| 256 | 12 | 1 | 11108% | 131% |

## VII. COMPUTATIONAL COMPLEXITY

We now consider the generic problem $\mathbb{P}$ of designing a universal ML-AM, which is solved by an algorithm alg if and only if given fixed $m$ and $n$, and for any set $\mathcal{S}$ of $m$ elements over $\mathcal{A}^n$ as input, alg$[\mathcal{S}]$ is a ML-AM for $\mathcal{S}$.

We are interested in the complexity $\Phi_{ret}$ of the retrieval process. Denote by $\Phi(\text{alg}[\mathcal{S}], \overline{w})$ the complexity of the operation of retrieving the unique $w \in \mathcal{S}$, if it is well defined, such that $d(w, \overline{w}) = 0$. Then

$$\Phi_{ret} = \min_{\text{alg solving } \mathbb{P}} \left[\max_{\mathcal{S}:|\mathcal{S}|=m} \left[\max_{\overline{w}:|\mathcal{S}(\overline{w})|=1} \Phi(\text{alg}[\mathcal{S}], \overline{w})\right]\right].$$

**Theorem 6.** *The retrieval complexity scales as*

$$\Phi_{ret} = \omega(n) \quad \text{operations}.$$

*Proof:* We obtain the lower bound by constructing an appropriate set $\mathcal{S}$. Specifically, consider sets of the form $\mathcal{S} = \{a^k b a^{n-k-1}, a, b \in \mathcal{A}\}$. Given a single erasure in the input $\overline{w}$ that happens to be the unique $b$ in the initial word, it is clear that $\mathcal{S}(\overline{w})$ is a singleton. To find the correct answer, the algorithm must read at least the erased symbol. It follows that the algorithm must read at least all symbols in the input and thus has complexity at least $n$. This reasoning can be extended to any set that contains $\mathcal{S}$ and not $\{a^n\}$, and thus to any size of set $n \leq m < |\mathcal{A}|^n$. ∎

Reciprocally one can achieve this complexity on random sets, at the cost of a large amount of memory. Consider the

following algorithm TBA (for Trie-Based Algorithm), based on tries [23], [24]:

Receiving a set $\mathcal{S}$: **begin**
    **for** *each permutation $\sigma$ of $[1;n]$* **do**
        Create the trie $\mathcal{T}_\sigma$ of $\mathcal{S}_\sigma = \{\sigma(w), w \in \mathcal{S}\}$.
    **end**
**end**

Retrieving a word from its partially erased version $\overline{w}$: **begin**
    Set j to 1
    Set $\sigma$ to $[1; 2; 3; \ldots; n]$
    **for** *i from 1 to n* **do**
        **if** $\overline{w}(j) = \perp$ **then**
            Swap $\sigma(j)$ and $\sigma(i)$
            Increment i
        **end**
    **end**
    Use $\mathcal{T}_{\sigma(\overline{w})}$ up to the first $\perp$ symbol
    Choose any path in $\mathcal{T}_{\sigma(\overline{w})}$ from there to the end
**end**

**Algorithm 1:** Universal ML-AM algorithm that reaches the minimum complexity for retrieving: $\Theta(n)$.

Table II compares the complexities of HNN, GBNN and TBA. Note that, to be fair, the complexity of the storing process has also been added to the table.

## VIII. CONCLUSION

We derive bounds on performance, memory, and computational complexity of maximum likelihood associative memories. When the number of messages $m$ in the dataset is small compared to the number of possible messages, a brute force approach has the best performance and optimal memory usage. In similar conditions, TBA offers the best performance and best computational complexity. However, brute force has a dramatic computational complexity and TBA has space complexity which is exponentially larger than the other methods.

When all three dimensions must be considered jointly, approaches based on neural networks offer an interesting tradeoff.

TABLE II
COMPARISON OF THE COMPLEXITIES OF HNN, GBNN WHEN CONSIDERING A CONSTANT NUMBER OF ITERATIONS AND THAT OF TBA.

| Algorithm | Complexity for retrieving | Complexity for storing |
|---|---|---|
| HNN | $\Theta(\binom{n2^{|\mathcal{A}|}}{2})$ | $O(mn^2)$ |
| GBNN | $\Theta(\binom{n}{2}|\mathcal{A}|^2)$ | $O(mn^2)$ |
| TBA | $\Theta(n)$ | $\Omega(mn!)$ |